\documentclass[twocolumn,showpacs,preprintnumbers,amsmath,%
amssymb,prb,superscriptaddress]{revtex4}

\usepackage{graphicx}
\usepackage{color}

\usepackage[pdfsubject={electronic structure},
pdfauthor={H. Gretarsson},pdftitle={RIXS KFe2Se2},%
pdfpagetransition=Glitter,colorlinks=true,plainpages=false,linkcolor=blue,urlcolor=blue,citecolor=blue,pdfpagemode=UseNone,pdfstartview=FitBH]{hyperref}

\def\HG#1 {\emph{\color{blue}#1}}
\begin{document}

\title{Resonant inelastic x-ray scattering study of electronic excitations in insulating K$_{0.83}$Fe$_{1.53}$Se$_2$}

\author{H.~Gretarsson}
\email{h.gretarsson@fkf.mpg.de}
\affiliation{Max-Planck-Institut f\"{u}r Festk\"{o}rperforschung, Heisenbergstr. 1, D-70569 Stuttgart, Germany}
\affiliation{Department of Physics, University of Toronto, 60
St.~George St., Toronto, ON, Canada M5S 1A7}
\author{T. Nomura}
\affiliation{Japan Atomic Energy Agency, SPring-8, 1-1-1 Kouto, Sayo, Hyogo 679-5148, Japan}
\author{I.~Jarrige}
\affiliation{Photon Sciences Directorate, Brookhaven National Laboratory, Upton, New York, 11973, USA}
\author{A.~Lupascu}
\affiliation{Department of Physics, University of Toronto, 60
St.~George St., Toronto, ON, Canada M5S 1A7}
\author{M.~H. Upton}
\author{Jungho~Kim}
\author{D.~Casa}
\author{T.~Gog}
\affiliation{Advanced Photon Source, Argonne National Laboratory,
Argonne, Illinois 60439, USA}
\author{R.~H.~Yuan}
\author{Z.~G.~Chen}
\affiliation{Beijing National Laboratory for Condensed Matter Physics, Institute of Physics, Chinese Academy of Sciences, Beijing 100190, China}
\author{N.-L.~Wang}
\affiliation{International Center for Quantum Materials, School of Physics, Peking University, Beijing 100871, China}
\author{Young-June Kim}
\email{yjkim@physics.utoronto.ca}
\affiliation{Department of
Physics, University of Toronto, 60 St.~George St., Toronto, ON,
Canada M5S 1A7}

\date{\today}

\begin{abstract}

We report an Fe $K$-edge resonant inelastic X-ray scattering (RIXS) study of K$_{0.83}$Fe$_{1.53}$Se$_2$. This material is an insulator, unlike many parent compounds of iron-based superconductors. We found a sharp excitation around 1 eV, which is resonantly enhanced when the incident photon energy is tuned near the pre-edge region of the absorption spectrum. The spectral weight and line shape of this excitation exhibit clear momentum dependence. In addition, we observe momentum-independent broad interband transitions at higher excitation energy of 3-7 eV. Calculations based on a 70 band $dp$ orbital model,  using a moderate $U_{\rm eff}\approx 2.5$ eV, indicate that the $\sim$1 eV feature originates from the correlated Fe 3$d$ electrons, with a dominant $d_{xz}$ and $d_{yz}$ orbital character. We find that a moderate $U_{\rm eff}$ yields a satisfying agreement with the experimental spectra, suggesting that the electron correlations in the insulating and metallic iron based superconductors are comparable.
\end{abstract}

\maketitle

\noindent

\section{Introduction}

Resonant Inelastic X-ray Scattering (RIXS) has emerged as a powerful technique to study momentum dependent magnetic- and charge-excitations in correlated materials.\cite{Luuk11} In particular  a significant insight into the dynamics of high-temperature superconducting cuprates has been gained (for a recent review see Ref.~\citenum{Dean2014}). While Cu $L_3$-edge RIXS (soft RIXS) is well suited for studying  magnetic excitations in cuprates,\cite{Braicovich09} Cu $K$-edge RIXS (hard RIXS) is useful for understanding their charge dynamics.\cite{Hasan00,YJKim2002} However in the case of iron based superconductors, experiments have been limited in both regimes.

Unlike cuprates, Fe in pnictides/chalcogenides is in a tetrahedral environment, where on-site mixing between Fe $4p$ and $3d$ orbitals is allowed. This mixing allows strong $4p \rightarrow 1s$ type fluorescence emission that dominates most RIXS spectra,\cite{Jarrige_XES} which is not the case in cuprates where Cu is in octahedral environments. Observation of charge or spin excitations in these Fe compounds with RIXS is thus very challenging, as the RIXS features are usually much weaker than the fluorescence line. In their Fe $K$-edge RIXS experiment, Jarrige and coworkers circumvented this problem by utilizing resonance at a much higher incident energy; here the contribution from the fluorescence line was much smaller. Comparison of their results with {\em ab initio} calculation supported a moderate Coulomb repulsion of $U_{\rm eff} \approx 2.5$ eV  in the parent PrFeAsO compound.\cite{Jarrige_RIXS} At the Fe $L_3$-edge where spectra are also dominated by a strong fluorescence signal,\cite{Yang2009} a recent RIXS experiment succeeded in observing paramagnons in (Ba$_{1-x}$K$_{x}$)Fe$_2$As$_2$,\cite{FeL3_RIXS} in a similar fashion to a recent study on cuprates.\cite{Keimer11} Although it was difficult to extract momentum dependence of the observed excitations, due to extremely low count rate, these results are encouraging. Moreover they demonstrate that both Fe  $K$- and $L_3$-edge RIXS can provide us with new insight into the physics of the iron based superconductors, just like for cuprates.

Among iron based superconductors, alkali metal iron selenides present an interesting case for RIXS studies. These compounds, with a generic chemical composition A$_x$Fe$_{(2-y)}$Se$_2$ (A=K, Cs and Rb),\cite{Guo_KFe2Se2:2010, Wang_RbFe2Se2:2011,Krzton_CsFe2Se2:2011,Dagotto_RMP:2013} are in many ways unique among  isostructural iron pnictides.\cite{Johnston2010} For instance, measurements of the Fermi surfaces, by angle-resolved photoemission spectroscopy (ARPES),\cite{Kuroki_PRL:2008,Feng_NatMat:2011,Ding_PRL:2011,Zhou_PRL:2011} show that these systems exhibit no nesting properties of electron- and hole-like Fermi surfaces, which is incompatible with the Fermi surface driven spin-fluctuation mediated superconductivity.\cite{Mazin_PRL:2008,Chubukov_PRB:2008,FaWang_PRL:2009} However, one of the unresolved issues in these materials is the intrinsic phase separation. There is not a consensus on whether the parent compound  is insulating,\cite{FangEPL_PhaseDiagram:2011} semiconducting,\cite{Chen2011} or even metallic.\cite{WeiPRL_STM:2012} A phase separation \cite{Bianconi:2011,Deisenhofer:2011,Amato:2011} has been suggested to explain these observations. Interestingly, the problem of phase separation, which exists in the superconducting materials (insulating/magnetic and metallic domains),\cite{Bianconi:2011,Deisenhofer:2011,Amato:2011} seems to disappear in the insulating samples which can be obtained  for a particular value of $y$ (e.g. K$_{0.83}$Fe$_{1.53}$Se$_2$). \cite{FangEPL_PhaseDiagram:2011,WeiBao_KFe2Se2_Neutron:2011} Moreover, for that same value of $y$, a novel blocked antiferromagnetic (AFM) order with a large magnetic moment is found. \cite{WeiBao_KFe2Se2_Neutron:2011} These observations suggest that K$_{0.83}$Fe$_{1.53}$Se$_2$ could give us an opportunity to study an insulating compound with Fe $K$-edge RIXS.

In this work, we report an Fe $K$-edge RIXS study of insulating K$_{0.83}$Fe$_{1.53}$Se$_2$. We find a sharp $dd$-excitation around 1 eV, whose spectral weight and line shape change as momentum transfer is varied. A broad Fe $4p$ interband transition is also observed at much higher energy of 3-7~eV, which is momentum-independent. Calculations based on 70 band $dp$ model, using a moderate $U_{\rm eff}\approx 2.5$ eV, can capture the $\sim\!1$ eV feature, which is found to have a dominant $d_{xz}$ and $d_{yz}$ orbital character,  while a 102 band orbital model, which takes into account the Fe $4p$ states as well, is needed to describe the interband transition. Our findings suggest that although  K$_{0.83}$Fe$_{1.53}$Se$_2$ is an insulator, it has a $U_{\rm{eff}}$ comparable to that of metallic iron pnictides. In addition, we discuss the observed behavior of $dd$-excitations in  K$_{0.83}$Fe$_{1.53}$Se$_2$ in comparison with a similar excitation in PrFeAsO \cite{Jarrige_RIXS} and cuprates.

\section{Experimental Details}

The RIXS experiment was carried out at the Advanced Photon Source using the 30ID MERIX spectrometer.
A spherical (1~m radius) diced Ge(620) analyzer was used and an
overall energy resolution of 230 meV (FWHM) was obtained. The same experimental configuration was used in Ref. \citenum{Jarrige_RIXS}.
The energy calibration was based on the absorption spectrum
through a thin Fe foil. Most of the measurements were carried out in a horizontal scattering geometry near {\bf Q} = (0 0 11) for which the scattering angle 2$\theta$ was close to 90$^\circ$, in order to minimize the elastic background intensity. The sample was freshly cleaved just before being mounted on a closed-cycle refrigerator. Details of the growth and the characterization of single-crystal samples were reported earlier.\cite{KFe2Se2_Optics:2011} In-plane dc resistivity and magnetic susceptibility data confirm that our sample is AFM insulator. Throughout this paper, we use for simplicity the tetragonal high temperature $I\rm{4/mmm}$ unit cell with two Fe atoms per lattice point ($a$ = $b$ = 3.8 $\rm{\AA}$ and $c$ = 13.6 $\rm{\AA}$). In this notation the observed Fe vacancy order and the blocked AFM order, associated with the $\sqrt{5}\times\sqrt{5}\times 1$ tetragonal $I\rm{4/m}$ unit cell,  appear at  {\bf Q}$_S$ = (0.2, 0.6, 0) and {\bf Q}$_M$ = (0.4, 0.2, 0), respectively.\cite{Ye_PRL:2011} The high x-ray energy used allows us to keep the rotation of the sample within 10$^\circ$ when measuring over the whole Brillouin zone, therefore minimizing any matrix element effect on our spectra.


\begin{figure}[htb]
\includegraphics[width=0.9\columnwidth]{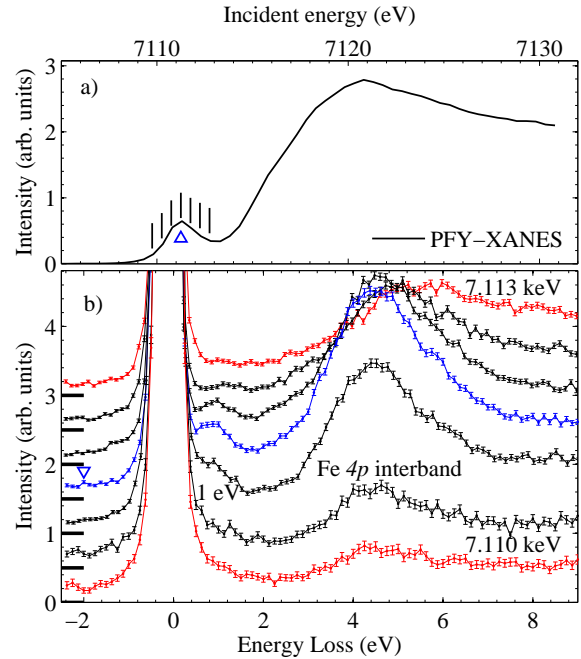}
\caption{\label{fig01}(Color online) (a) Fe $K$-edge PFY-XANES taken by monitoring the intensity of K$\beta_{1,3}$ emission line $(3p \rightarrow 1s)$ as a function of incident energy ($E_i$). Ticks represent the incident energies used for our energy dependence in (b) and the blue triangle our resonance energy. (b) RIXS spectra as a function of the incident energy. Measurements were carried out at {\bf Q} = (0.5 0.5 10.75) at T = 15 K. The spectra have been shifted vertically for clarity (horizontal ticks).}
\end{figure}

\section{Experimental Results}
\subsection{Incident Energy dependence}
The Fe $K$-edge x-ray absorption near-edge spectra (XANES) taken in the partial fluorescence yield mode is shown in  Fig.~\ref{fig01}(a). The spectra were obtained by monitoring the intensity of K$\beta_{1,3}$ emission line $(3p \rightarrow 1s)$ as a function of incident energy ($E_i$). Two distinct features are seen in the spectra, a sharp pre-edge peak around $E_i$~=~7.111 keV, corresponding to excitations of $1s$ electrons into the empty Fe $3d/4p$ states hybridized with Se $4p$ states, and the main edge around $E_i$~=~7.120 keV, corresponding to excitations into  mostly empty Fe $4p$ states.\cite{Saini_XAS}

\begin{figure}[htb]
\includegraphics[width=\columnwidth]{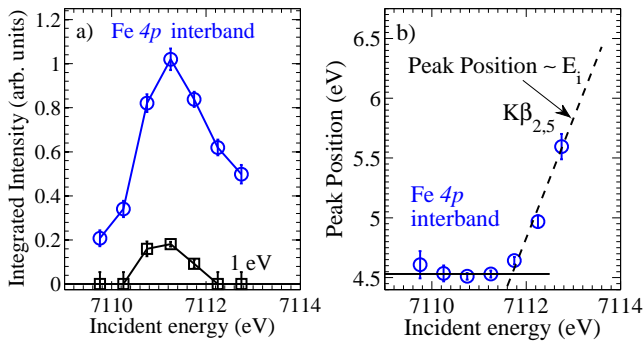}
\caption{\label{fig04}(Color online) (a)  The evolution of the integrated intensity of both the Fe $4p$ interband transition and  the 1 eV features around the pre-edge. (b) Peak position of the Fe~$4p$ interband transition as a function of $E_i$. Numbers were derived from fit (see text).}
\end{figure}

The incident energy dependence of the RIXS spectra for energies around the pre-edge (as indicated by vertical lines in  Fig.~\ref{fig01}(a)) is plotted in Fig.~\ref{fig01}(b). A broad and strong inelastic feature is seen in the $\hbar\omega=$3-7 eV  range ($\hbar\omega$ = $E_i - E_f$ with $E_f$ as the energy of the outgoing x-ray). We assign this feature to an Fe $4p$ interband transition. Such a transition is possible since the absence of a center of symmetry at the tetrahedral Fe site allows a significant on-site mixing between Fe $3d$ and $4p$ orbitals to occur, pushing some of the Fe $4p$ states below the Fermi level. Therefore one can excite electrons from predominantly $4p$ band to the hybridized $3d$-$4p$ band just above the Fermi level. In between the Fe $4p$ interband transition and the elastic peak ($\hbar\omega$ = 0) we observe a clear shoulder-like RIXS feature around $\hbar\omega \approx$ 1~eV. Their intensity and position dependence on incident energy, $E_i$, was investigated. We fitted two spectral features using two Gaussian functions of fixed widths and a linear background to account for the K$\beta_{1,3}$ emission line at higher energy loss. In Fig. \ref{fig04} (a) we plot the integrated intensity of the 1 eV feature and the Fe $4p$ interband transition as a function of $E_i$; in both cases a large resonance enhancement near $E_i$ = 7.111~keV is observed.  At higher incident energies the Fe $4p$ interband transition loses intensity and evolves into the K$\beta_{2,5}$ emission line (includes $4p\rightarrow 1s$ transition). Since emission occurs at fixed outgoing photon energy ($E_f$), the peak position as a function of energy transfer ($\hbar \omega$) is proportional to $E_i$, following a linear dashed line as shown in Fig.~\ref{fig04}(b). The 1~eV peak is however only visible around the resonant incident energy.

 \begin{figure}[htb]
\includegraphics[width=\columnwidth]{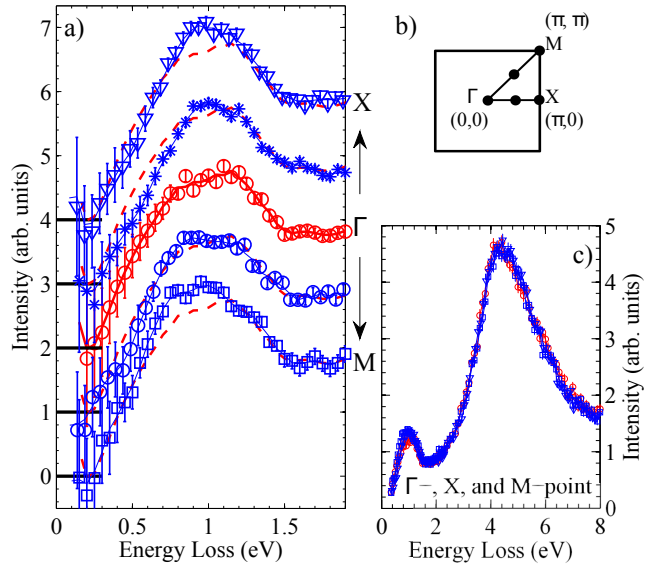}
\caption{\label{fig02}(Color online) (a) Momentum dependence of the low energy RIXS spectra of K$_{0.83}$Fe$_{1.53}$Se$_2$  obtained at $T=15$~K. Contribution from the elastic line has been subtracted. The spectra have been shifted vertically for clarity (horizontal ticks) and the solid lines are three-point smoothed spectra.
Superimposed as dashed red line is the smoothed spectrum taken at $\Gamma$-point. In (b) a schematic diagram of the (H K 0) reciprocal space is shown. The Brillouin zone (BZ) corresponding to the tetragonal unit cell (with two Fe per lattice point) is shown as a solid line. The filled circles are the points where RIXS spectra in (a) are taken. (c) Wide range RIXS spectrum at $\Gamma$-, $X$-, and $M$-points. }
\end{figure}

\subsection{Momentum dependence}
 Further insight into the nature of the 1 eV feature can be gained by measuring its momentum dependence. In Fig.~\ref{fig02} (a) we plot the low energy part of the RIXS spectra of K$_{0.83}$Fe$_{1.53}$Se$_2$ after subtracting contributions from the elastic line. The elastic line background was obtained by measuring the off-resonance spectrum at $E_i=7.107$ keV, as done previously in the study of two-magnon excitations in cuprates.\cite{Ellis2010} Each spectrum was normalized by the Fe  $4p$ interband transition intensity, which does not show any variations with {\bf q}. This is evident from Fig. \ref{fig02} (c) where we compare spectra at $\Gamma$-, $X$-, and $M$-points. Such a lack of momentum dependence allows us to use the Fe $4p$ interband transition to normalize each spectrum in Fig.~\ref{fig02} (a). The momentum depedence of the 1~eV feature was measured along the high-symmetry directions shown in the Brillouin zones (BZ) in Fig.~\ref{fig02} (b), where $\Gamma$ is {\bf Q} = (0 0 11.5), $X$ is {\bf Q} = (0.5 0 11.25), and $M$ is {\bf Q} = (0.5 0.5 11.3). Although overall intensity and peak position do not change drastically, one can clearly observe the difference between the $\Gamma$ point spectrum (middle) with the zone boundary spectra in either direction (top or bottom). At the $\Gamma$ position, the spectral weight is at higher energy side, with maximum intensity occurring near 1.1 eV. As you move away from $\Gamma$, the center of mass of the feature shifts to lower energy and the peak seems to become sharper with maximum intensity near 0.9~eV. Since we only have data for a limited number of {\bf q}-points and the spectral variation is small, extracting any dispersion relation is difficult. To emphasize the momentum dependence we have superimposed the $\Gamma$-point data as dashed red lines for each spectrum. We can clearly observe enhanced intensity around 0.9 eV for both zone-boundary $M$- and $X$-points; scans along $\Gamma-M$ and $\Gamma-X$ exhibit similar variations.

\section{Theoretical Calculations}
In order to understand the origin of the observed spectral features and their momentum dependence, we performed first-principles electronic structure calculations for $\sqrt{5}\times\sqrt{5}$ Fe vacancy ordered  K$_{0.8}$Fe$_{1.6}$Se$_2$ using the WIEN2k code.\cite{Code} By fitting the band structure to a tight-binding model, using the wannier90 code,\cite{Mostofi2008} the RIXS spectra could be explicitly calculated. Short description on the calculation method is found in Appendix \ref{app:RIXStheory}, for detailed description see Refs.~\citenum{Nomura2005,Takahashi2007}. We used two different tight-binding models. The first is a 102-orbital model, which includes Fe $3d$, Fe $4s$, Fe $4p$, and Se $4p$ orbitals in order to reproduce the density of states (DOS) near the Fermi level ($E_F$) as precisely as possible. The second, a 70-orbital model, inluding only Fe $3d$ and Se $4p$ orbitals, is used to calculate the low-energy RIXS spectral weight, which has a strong contribution from the correlated Fe $3d$ electrons. We note that since the band structure calculations overestimate the bandwidth of the states in the energy range  -2~eV~$\!\!<\!\!E\!<\!\!$~1~eV, a renormalization factor of $\sim\!2$ was needed. This factor is in fact consistent with the 2-2.5 bandwidth renormalization factor previously used to describe ARPES results with LDA calculations. \cite{DingARPES_Ba122:2011,DingARPES_K122:2011}

\subsection{102-orbital model: Fe $4p$ interband transition}

\begin{figure}[htb]
\includegraphics[trim=0.0cm 0cm -2.0cm 0cm, width=\columnwidth]{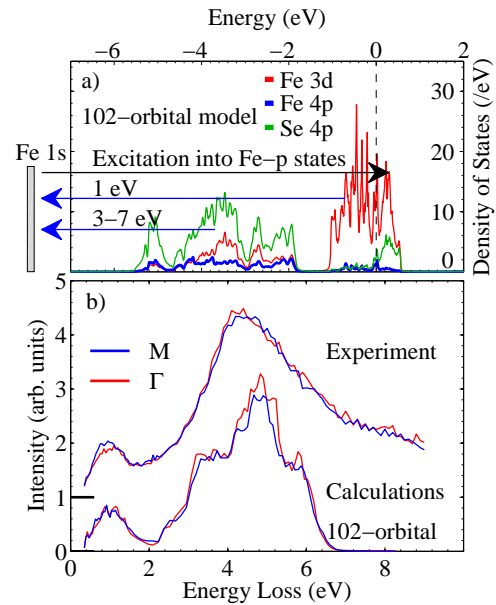}
\caption{\label{fig03}(Color online) (a) Fe 3$d$/4$p$ and Se 4$p$ partial density of state in the nonmagnetic state.
Data was obtained by the 102-orbital model (see the text). The Fermi level is set to 0~eV.
Arrows indicate the considered RIXS processes.
(b) Calculated momentum-resolved ($\Gamma$ and $M$-point) RIXS spectra
using the 102-orbital model. For comparison we also include the experimental data. The spectra have been shifted vertically for clarity (horizontal ticks). }
\end{figure}

In Fig. \ref{fig03} (a) the Fe 3$d$/4$p$ and Se 4$p$ DOS from the 102-orbital model is plotted.
Since the Fe $4p$ partial DOS extends across $E_F$, the Fe $4p$ states below $E_F$
should contribute significantly to the RIXS spectra.
The following excitation process is considered: a $1s$ core-electron is promoted,
through the absorption of a photon (black arrow), into the empty Fe $4p$ states
and subsequently an Fe $3d$ or $4p$ electron below $E_F$ fills the $1s$ core-hole (blue arrows),
leaving behind an excited state. In Fig. \ref{fig03} (b) we plot this calculated RIXS spectrum as obtained from the 102-orbital model at the $\Gamma$- and $M$-point. To account for the observed optical gap we have shifted the calculated spectrum by 350 meV towards higher energy loss.
The overall agreement between the experimental results and our calculated RIXS spectrum is very good. We can see that the two separate excitations at 1 eV and 3-7 eV are largely accounted for by the occupied density of states. We note that, since the Fe 4$p$ states (intermediate state) have stronger hybridization with neighboring Se or Fe 4$p$ states than adjacent Fe 3$d$ states, the 3-7 eV excitation becomes more pronounced than the 1 eV excitation.
However, despite the good overall agreement, the observed momentum dependence of the 1 eV excitation cannot be accounted for by the current 102-orbital model, which ignores electron correlation effect.

\subsection{70-orbital model: RIXS}

Since the low-energy RIXS spectral features are expected to originate from Fe $3d$ and Se $4p$, we carried out a separate calculation using the 70-orbital model which only focuses on these orbitals. This reduction in number of orbitals allows us to include the electron correlation, which is too computationally taxing for the 102-orbital model.
In the 70-orbital calculation, processes to screen the $1s$ core-hole are included and
the random-phase approximation is used to account for Fe $3d$ electron correlations.\cite{NomuraRIXS:2014} The block-checkerboard AFM ordering \cite{WeiBao_KFe2Se2_Neutron:2011} is described within the Hartree-Fock approximation. \cite{Dagotto_PRB:2011} The on-site Coulomb integrals are included as $U$ (intra-orbital), $U^\prime=0.6 \times U$ (inter-orbital), $ J = J^\prime = 0.2 \times U$ (Hund's coupling).
For $U=5$ eV, an energy gap of about 540 meV at $E_F$
and an ordered moment of 3.5 $\mu_B$ are obtained, which is in agreement with neutron scattering\cite{WeiBao_KFe2Se2_Neutron:2011}
and optical conductivity.\cite{KFe2Se2_Optics:2011} Recall that the bandwidth was normalized by a factor of $\sim\!2$ in our band structure calculation to be consistent with ARPES data. \cite{DingARPES_Ba122:2011,DingARPES_K122:2011} In order to keep the physically meaningful ratio $U/W$ unchanged, the experimentally relevant energy scale is $U_{\rm eff}\sim U/2$.

\begin{figure}[htb]
\includegraphics[trim=0.0cm 0cm -0.5cm 0cm,width=0.80\columnwidth]{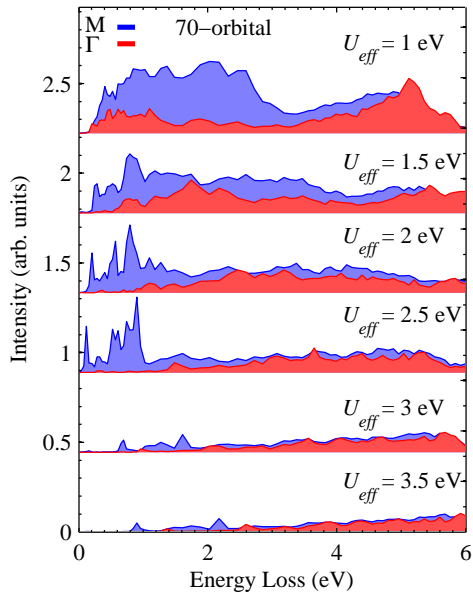}
\caption{\label{fig05}(Color online)  Contribution from Fe $3d$/Se 4$p$ states to the RIXS spectrum shown over wide energy range, as derived from the 70-orbital model. Calculations are presented for different transferred momentum and $U_{\rm eff}$. Note that the intensity scale used in this figure is the same as for Fig. \ref{fig03}(b).  }
\end{figure}

We start by investigating the effect of Coulomb repulsion, $U_{\rm eff}$, on our calculated RIXS spectrum. In Fig. \ref{fig05} we plot the contribution from Fe $3d$/Se 4$p$ states to the RIXS spectrum over wide energy range as derived from the 70-orbital model. Calculations are shown for both $\Gamma$-point (red) and $M$-point (blue) for 1 eV $\leq U_{\rm eff} \leq$ 3.5 eV. Overall RIXS spectrum changes significantly as $U_{\rm eff}$ is varied. As $U_{\rm eff}$ increases from 1 eV to 2.5 eV we notice that the spectrum above 1~eV, which arises from Fe-Se $pd$ charge-transfer (CT), is suppressed. In particular, the large difference between the $\Gamma$- and M-point spectra present for $U_{\rm eff}=1$ eV more or less disappears for $U_{\rm eff}=2.5$ eV. By increasing $U_{\rm eff}$ above 2.5 eV, the momentum dependence of the spectrum below 1~eV is also greatly suppressed,  resulting in almost quenched momentum dependence in the entire energy range. Based on these calculations and our results in Fig. \ref{fig02}(a), where variations with momentum were limited to the spectrum below 1 eV, we find that $U_{\rm eff}\sim 2-2.5$ eV gives satisfying agreement with our results. We would like to point out that this $U_{\rm eff}$ value is in line with values obtained for a realistic blocked AFM state in a five orbital Hubbard model.\cite{Dagotto_PRB:2011} This suggests that even in this insulating compound, a moderate Coulomb repulsion similar to the iron pnictides is sufficient.\cite{Jarrige_RIXS}

\begin{figure}[htb]
\includegraphics[width=1\columnwidth]{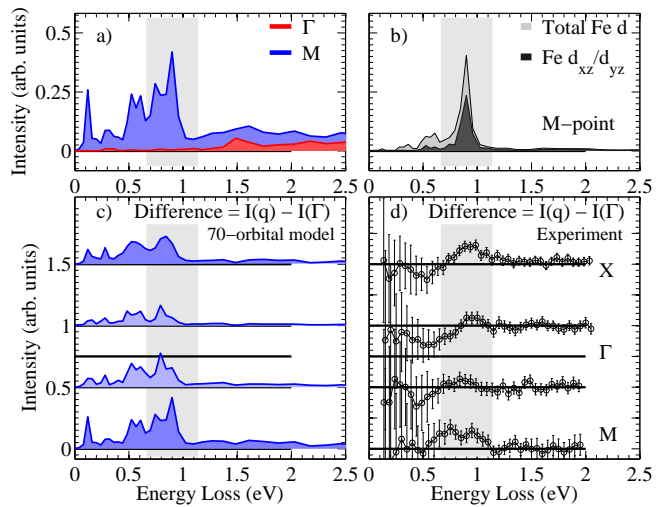}
\caption{\label{fig06}(Color online)  (a) Contribution from Fe $3d$/Se 4$p$ states at different transferred momentum. (b) Orbital resolved contribution of the Fe $3d$ states at $M$-point. The lighter-shaded area represents the total contribution from the Fe $3d$ states while the darker-shaded area a transition into only the $d_{xz}$ and $d_{yz}$ orbitals. The plot demonstrates that the 0.9 eV feature consists mostly of excitation into the $d_{xz}$ and $d_{yz}$ orbitals. (c) Difference between the calculated RIXS spectra at momentum {\bf q} and $\Gamma$-point. (d) Difference between the experimental RIXS spectra from Fig. \ref{fig02}(a) and the $\Gamma$-point spectrum (dashed red line in Fig. \ref{fig02}(a)) .}
\end{figure}

In order to examine the momentum dependence in detail, the low-energy region of the calculated RIXS spectra from the 70-orbital model with $U_{\rm eff}=2.5$ eV are plotted in Fig.~\ref{fig06}(a)-(c). In Fig. \ref{fig06}(a), contributions from the Fe $3d$ and Se $4p$ states are highlighted. While these states have rather limited contributions below 1 eV at the $\Gamma$-point,  a large increase of spectral weight below 1 eV is observed at the  $M$-point. By looking at the orbital resolved contribution of the Fe $3d$ states at the $M$-point in Fig. \ref{fig06} (b) we notice that the 1 eV feature originates mostly from the Fe $3d$ states (lighter-shaded area), with the largest contribution from a transition involving the $d_{xz}$ and $d_{yz}$ orbitals (darker-shaded area): $d_{yz} \rightarrow d_{yz}$, $d_{xz} \rightarrow d_{xz}$, $d_{x^2-y^2} \rightarrow d_{yz}$, and $d_{x^2-y^2} \rightarrow d_{xz}$.

In Fig. \ref{fig06} (c) the low energy RIXS spectrum, as obtained from the 70-orbital model, is shown for momentum transfers along the $\Gamma$-$X$ and $\Gamma$-$M$ directions. In order to emphasize the change with respect to the spectrum at $\Gamma$-point, the $\Gamma$-point spectrum has been subtracted from each spectrum. In Fig. \ref{fig06} (d), we plot the experimental data in the same manner. Here we have subtracted the dashed red line ($\Gamma$-point) from the respective RIXS spectra in Fig. \ref{fig02} (a). The calculation shows the spectral weight around 1~eV (gray vertical bar) increases significantly as we move away from $\Gamma$-point. Similar momentum dependence, although weaker, is observed in the experimental data. Therefore, we conclude that including moderate correlation energy $U_{\rm eff}$=2.5~eV in the 70-orbital calculation allows us to describe the observed momentum dependence of the 1~eV feature in our RIXS data.

\section{Discussion}

To summarize our results, we find two spectral features, a sharp 1~eV peak and a broad feature around 3-7 eV, in our Fe $K$-edge RIXS investigation of K$_{0.83}$Fe$_{1.53}$Se$_2$, an insulating iron chalcogenide. The observed low-energy feature exhibits a weak {\bf q}-dependence, while the broad high energy peak is {\bf q}-independent. Overall energy positions and intensities of these two excitations can be captured using the 102-orbital model, which include Fe $3d$, $4s$, and $4p$, and Se $4p$ states. However, the 102-orbital model, which does not take into account electron correlation, can not describe the {\bf q}-dependence of the 1~eV feature. By using instead the 70-orbital model, which focuses on the Fe $3d$ and Se $4p$ states and takes into account electron correlation, a satisfactory description of the 1~eV feature is achieved using $U_{\rm eff}=2.5$ eV.

Poor experimental statistics prevented us from extracting quantitative dispersion relation of the 1 eV feature. Given that there seems to be multiple transitions contributing to this feature from our calculation, the momentum dependence could be due to changing spectral weight of these transitions. However, we point out that since the momentum dependent 1~eV feature consists mostly of excitations involving the $d_{xz}$ and $d_{yz}$ orbitals a  sizable orbital dependent correlations could exist in the insulating K$_{0.83}$Fe$_{1.53}$Se$_2$. These are the same orbitals that are believed to give rise to the observed nematicity in the 122 FeAs.\cite{Fisher2010, Shen2011, Kasahara2012}
The $d_{xy}$ orbital, which was found to go through metal-insulator (Mott) transition in the superconducting 122 FeSe samples,\cite{WangNatComm_OSMI:2014,YiPRl_OSMI:2013} does not make a significant contribution to the 1~eV feature. Our observation seems to lend support to theoretical models that consider both orbital- and spin-fluctuation to contribute to the strong pairing interaction needed for superconductivity.\cite{Kontani_PRB:2013}

We would like to mention that Chen and coworkers observed a sharp double-peak structure below 1~eV in their optical conductivity measurements of K$_{0.83}$Fe$_{1.53}$Se$_2$, which were attributed to arising from the particular magnetic structure.\cite{KFe2Se2_Optics:2011} Each Fe spin has two types of neighbors in this block antiferromagnetic ordered state: one aligned ferromagnetically and the other antiferromagnetically with the original spin.
Intersite $dd$-excitations to these different spin arrangements have different energies, and the two peak structure in optical spectrum arises from these two transitions. However, one cannot directly compare RIXS data with these optical spectroscopy observations due to the difference in the response function these techniques are probing. RIXS follows symmetry selection rules of Raman scattering, and mostly sensitive to {\em intrasite} $dd$-excitation, rather than {\em intersite} $dd$-excitation. Perhaps this is the reason why we observe only one peak in our RIXS data, since local $dd$-excitation should always satisfy the spin-selection rule.

Next, we compare the current results with previous Fe $K$-edge RIXS experiment on PrFeAsO.\cite{Jarrige_RIXS} Overall spectra bear many similarities, but there are some differences. First, in both systems a non-dispersive excitation at 3-5 eV was found. The Fe $4p$ interband transition energy is also higher in K$_{0.83}$Fe$_{1.53}$Se$_2$ by about 600 meV, when compared to PrFeAsO.\cite{Jarrige_Unpublished} In addition, the higher incident energy used in the study of PrFeAsO seems to suppress the intensity of the Fe $4p$ interband transitions. As a result, the 3-5 eV feature in PrFeAsO was mostly associated with CT. In K$_{0.83}$Fe$_{1.53}$Se$_2$ we also find in our 70-orbital model a non dispersive Fe-Se $pd$ CT excitation in the same energy interval (3-7 eV) as the Fe $4p$ interband transition. However, the CT excitation could not be resolved in our RIXS spectrum and is instead hidden under the strong Fe $4p$ interband transition. Secondly, although in both K$_{0.83}$Fe$_{1.53}$Se$_2$ and PrFeAsO a low energy $dd$-excitation with a dominant $d_{yz}$ and $d_{xz}$ orbital character is found, their behaviour with momentum are quite different. Whereas in PrFeAsO the $dd$-excitation is present at the $\Gamma$-point and disperses towards higher energy (bandwidth $\sim 0.4$ eV) with increasing in-plane momentum, the 1~eV feature in K$_{0.83}$Fe$_{1.53}$Se$_2$ is much weaker at the $\Gamma$-point and exhibits very small dispersion when in-plane momentum is increased. Since the  dispersion of the $dd$-excitation in PrFeAsO was associated with local magnetic correlations, reflecting the collinear antiferromagnetic order, we speculate that the lack of much momentum dependence of the 1~eV feature might be influenced by the different magnetic ordering pattern.

Finally, it is quite illuminating to compare the RIXS spectra in Fe based superconductors and cuprates. The most prominent difference between the RIXS spectra of these two families of compounds is the difference in scattering cross-section. Cuprate RIXS spectra usually show prominent and well-defined charge-transfer excitations between 2-7 eV, while iron based materials do not show such a well defined feature. This contrast can be ascribed to the difference in the hybridization of the respective $d$-orbitals. In cuprates the spatial overlap between Cu $d_{x^2-y^2}$ orbitals and oxygen $p_{x,y}$ orbitals is quite large, and the hybridization is sizable. In the Cu $K$-edge process in cuprates the screening of the Cu $1s$ core hole is mainly provided by the charge transfer from the oxygen $p$ orbitals, resulting in salient features associated with the charge-transfer excitation in the spectra. On the other hand, in iron based superconductors Fe $d$ orbitals are in tetrahedral environment of pnictogen or chalcogen atoms, the hybridization between orbitals from the two sites is thus much smaller.
In addition, the tetrahedral environment allows on-site Fe $4p-3d$ hybridization. Therefore, there exist more channels to screen the $1s$ core hole in the Fe $K$-edge RIXS experiment, since Fe $d$, $p$, and Se $p$ states are all available for the screening. As a result, the spectral features are all quite weak and perhaps there are more contributions than one dominant charge-transfer excitation. Another consequence of the Fe $d-p$ hybridization is the large fluorescence signal in iron compounds. Since significant Fe $4p$ density of states are found near the Fermi level, fluorescence occurs at low energy transfer, which overlaps with the RIXS features. On the other hand, Cu $4p$ DOS is far away from the Fermi level in cuprates, and does not usually interfere with the RIXS spectra. For these two reasons, RIXS investigation of iron based superconductors are much more challenging. Depending on the specific spectral feature one wants to focus on, a judicious choice of incident energy should be considered in the experimental design process.

\section{Conclusions}
We have successfully measured charge excitations in the insulating  K$_{0.83}$Fe$_{1.53}$Se$_2$ using Fe $K$-edge resonant inelastic x-ray scattering (RIXS). Our key observation is the appearance of a sharp excitation around 1 eV when the incident energy is tuned to the pre-edge, as well as a broad spectral feature around 3-7 eV. This low energy $dd$-excitation shows clear momentum dependence of the spectral weight and line shape, while the high energy peak is due to Fe $4p$ interband transitions. Calculations based on 70 orbital model, using a moderate $U_{\rm eff} \approx\! 2.5$ eV, indicate that the 1~eV feature originates from the correlated Fe 3$d$ electrons with a dominant $d_{xz}$ and $d_{yz}$ orbital character, emphasizing the importance of those orbitals across different families of iron based superconductors. The momentum dependence of the $dd$-excitation in K$_{0.83}$Fe$_{1.53}$Se$_2$ is found to be quite different from a similar excitation in PrFeAsO,\cite{Jarrige_RIXS} we speculate that this most likely originates from their dissimilar magnetic ordering. Our results show that a moderate $U_{\rm eff}$ is in qualitative agreement with our RIXS spectrum, suggesting that comparable correlations can be found in the insulating and metallic iron based superconductors.

\acknowledgements{ Research at the U. of Toronto was supported by the Natural
Sciences and Engineering Research Council of Canada through
Discovery Grant and Research Tools and Instrument Grant. Use of the APS was supported by the U. S. DOE,
Office of Science, Office of BES, under Contract No. W-31-109-ENG-38.}

\appendix

\section{Formulas for the RIXS process}
\label{app:RIXStheory}

To discuss the RIXS process microscopically, we consider the following form of Hamiltonian: 

\begin{equation} 
H = H_{n.f.} + H_{1s} + H_{1s-3d} + H_x, 
\end{equation}

\noindent where $H_{1s}$ and $H_x$ describe the inner-shell $1s$ electrons and the dipole-transition by x-rays, respectively. 
$H_{n.f.}$ describes the electrons (Fe-$3d$, Se-$4p$, etc.) near the Fermi level (for this part we use the 102-orbital or 70-orbital model). 
$H_{1s-3d}$ is the Coulomb interaction between $1s$ and $3d$ electrons at Fe sites. 
For $1s$ electrons, we take completely localized $1s$ orbitals at each Fe site: 

\begin{equation}
H_{1s} = \sum_i \sum_{\sigma} \varepsilon_{1s}({\bf r}_i) s_{i\sigma}^{\dag} s_{i\sigma}, 
\end{equation}

\noindent where $\varepsilon_{1s}({\bf r}_i)$ is the one-particle energy of the Fe-$1s$ state, $s_{i\sigma}^{\dag}$ and $s_{i\sigma}$ are the creation 
and annihilation operators of $1s$ electrons with spin $\sigma$ at Fe site $i$, respectively. $H_x$ describes resonant $1s$-$4p$ dipole transition induced by x-rays: 

\begin{equation}
H_x =  \sum_{\mu = x,y,z} \sum_i \sum_{{\bf q}} \sum_{\sigma} 
w_{\mu}({\bf r}_i; {\bf q}, {\bf e}) \alpha_{{\bf q} {\bf e}} 
p_{i \mu \sigma}^{\dag} s_{i\sigma} + h.c., 
\end{equation} 

\noindent  where $p_{i \mu \sigma}^{\dag}$ is the creation operator of Fe $4p_{\mu}$ electron ($\mu=x,y,z$) at site $i$ with spin $\sigma$, 
and $\alpha_{{\bf q} {\bf e}}$ is the annihilation operator of a photon with momentum ${\bf q}$ and polarization ${\bf e}$.  
The matrix elements of  $w_{\mu}({\bf r}_i;{\bf q}, {\bf e})$ are given in the form: 

\begin{equation}
w_{\mu}({\bf r}_i; {\bf q}, {\bf e}) = - \frac{e}{m}\sqrt{\frac{2\pi}{|\bf q|}} 
e^{i {\bf q} \cdot {\bf r}_i } {\bf e} \cdot \langle 4p_{\mu} |{\bf p}|1s \rangle 
\propto e^{i{\bf q}\cdot {\bf r}_i}{\bf e} \cdot {\bf e}_\mu, 
\end{equation}

\noindent in natural units ($c = \hbar = 1$), where $e$ and $m$ are the charge and mass of an electron, respectively. ${\bf e}_\mu$'s are the orthonormal basis vectors. 
$H_{1s-3d}$ is given by 

\begin{equation}
H_{1s-3d} = \sum_i \sum_{\sigma\sigma'} V_{1s-3d}({\bf r}_i) 
s_{i\sigma}^{\dag}  s_{i\sigma} d_{i\sigma'}^{\dag} d_{i\sigma'}, 
\end{equation}

\noindent where $V_{1s-3d}({\bf r}_i)$ is the so-called core-hole potential at Fe site ${\bf r}_i$. 
In the present study, we take $V_{1s-3d}({\bf r}_i) = 2$eV 
and use the Born approximation. 

The RIXS formulae are: 
\begin{widetext}
\begin{eqnarray}
W_{4p}(q,q') &=& \frac{2 \pi}{N} \sum_{{\bf k}} \sum_{j_1j_2} 
n_{j_1}({\bf k}) [1-n_{j_2}({\bf k} + {\bf Q}) ]
\delta (\Omega + E_{j_1}({\bf k})- E_{j_2}({\bf k} + {\bf Q})) \nonumber \\
&& \times \biggl| \sum_{\mu,\mu' =x,y,z} 
\sum_i \sum_{\sigma} 
w_{\mu}({\bf r}_i; {\bf q}, {\bf e}) w_{\mu'}^*({\bf r}_i; {\bf q}', {\bf e}') 
\frac{u_{4p_{\mu}(i)\sigma, j_2}^*({\bf k}+{\bf Q})u_{4p_{\mu'}(i)\sigma, j_1}({\bf k})}
{\omega + \tilde{\varepsilon}_{1s}({\bf r}_i) - E_{j_2}({\bf k}+{\bf Q})} \biggr|^2 \nonumber\\
\label{Eq:W4p}
\end{eqnarray}
\end{widetext}

\noindent for Fe $4p$ inter-band processes~\cite{NomuraRIXS:2014}, and
\begin{widetext}
\begin{eqnarray}
W_{3d}(q,q') &=& \frac{2 \pi}{N} \sum_{\bf k} \sum_{j_1j_2} 
n_{j_1}({\bf k}) [1-n_{j_2}({\bf k} + {\bf Q}) ]
\delta (\Omega + E_{j_1}({\bf k})- E_{j_2}({\bf k} + {\bf Q})) \nonumber \\
&& \times \biggl| \sum_{\mu = x,y,z}  \sum_i 
w_{\mu}({\bf r}_i; {\bf q}, {\bf e}) w_{\mu}^*({\bf r}_i; {\bf q}', {\bf e}') 
\sum_{\ell_1\ell_2} \sum_{\sigma_1\sigma_2} \nonumber\\ 
&& V_{1s-3d}({\bf r}_i) \Lambda_{\ell_2\sigma_2, \ell_1\sigma_1}({\bf r}_i; Q) 
u_{\ell_2\sigma_2, j_2}^*({\bf k}_1+{\bf Q}) u_{\ell_1\sigma_1, j_1}({\bf k}) \nonumber\\
&&\times \int_{E_F}^{\infty} d \varepsilon \frac{ \rho_{4p_{\mu}(i)}(\varepsilon)}
{[\omega + \tilde{\varepsilon}_{1s}({\bf r}_i) - \varepsilon]
[\omega' + \tilde{\varepsilon}_{1s}({\bf r}_i) - \varepsilon]} \biggr|^2 
\label{Eq:W3d}
\end{eqnarray}
\end{widetext}
\noindent for Fe $1s$ core-hole screening processes~\cite{Nomura2005, Takahashi2007}, 
where $E_j({\bf k})$ is the energy of diagonalized band $j$, 
$n_j({\bf k})$ is the electron occupation number of band $j$ at  ${\bf k}$, 
$u_{..., j}({\bf k})$ are the diagonalization matrix elements, 
$\ell_{1,2}$ and $\sigma_{1,2}$ are orbital and spin indices for Fe $3d$ electrons. 
$q=(\omega, {\bf q})$  and $q'=(\omega', {\bf q}')$ are 
the four-vectors of incoming and outgoing photons, respectively, and 
$Q = q-q'=(\Omega, {\bf Q})$, where $\Omega = \omega - \omega'$ 
and ${\bf Q}={\bf q}-{\bf q}'$ are energy loss and momentum transfer, respectively. 
$4p_\mu(i)\sigma$ means the $4p_\mu$ state at Fe site ${\bf r}_i$ with spin $\sigma$. 
$\rho_{4p_{\mu}(i)}(\varepsilon)$ is the Fe $4p$ density of states, and is calculated 
by the band structure calculation. 
$\Lambda_{\ell_2\sigma_2,\ell_1\sigma_1}({\bf r}_i; Q)$ is a vertex function, 
which is calculated within RPA to take account of Fe $3d$ electron correlations. 
$\tilde{\varepsilon}_{1s}({\bf r}_i) \equiv \varepsilon_{1s}({\bf r}_i) + i \Gamma_{1s} $, 
where $\Gamma_{1s}$ is the damping rate of the $1s$ core hole 
and set to 0.8 eV in the present study. 
Summations in $i$ should be restricted only to eight Fe sites in the unit cell. 
We calculate eq.~(\ref{Eq:W4p}) using the 102-orbital model, 
and eq.~(\ref{Eq:W3d}) using the 70-orbital model, 
starting from the identical first-principles band structure. 
The incident photon energy $\omega$ is set to the pre-edge peak.


\begin{thebibliography}{48}
\expandafter\ifx\csname natexlab\endcsname\relax\def\natexlab#1{#1}\fi
\expandafter\ifx\csname bibnamefont\endcsname\relax
  \def\bibnamefont#1{#1}\fi
\expandafter\ifx\csname bibfnamefont\endcsname\relax
  \def\bibfnamefont#1{#1}\fi
\expandafter\ifx\csname citenamefont\endcsname\relax
  \def\citenamefont#1{#1}\fi
\expandafter\ifx\csname url\endcsname\relax
  \def\url#1{\texttt{#1}}\fi
\expandafter\ifx\csname urlprefix\endcsname\relax\def\urlprefix{URL }\fi
\providecommand{\bibinfo}[2]{#2}
\providecommand{\eprint}[2][]{\url{#2}}

\bibitem[{\citenamefont{Ament et~al.}(2011)\citenamefont{Ament, van Veenendaal,
  and van~den Brink}}]{Luuk11}
\bibinfo{author}{\bibfnamefont{L.~J.~P.} \bibnamefont{Ament}},
  \bibinfo{author}{\bibfnamefont{M.}~\bibnamefont{van Veenendaal}},
  \bibnamefont{and} \bibinfo{author}{\bibfnamefont{J.}~\bibnamefont{van~den
  Brink}}, \bibinfo{journal}{EPL (Europhysics Letters)}
  \textbf{\bibinfo{volume}{95}}, \bibinfo{pages}{27008} (\bibinfo{year}{2011}).

\bibitem[{\citenamefont{Dean}(2015)}]{Dean2014}
\bibinfo{author}{\bibfnamefont{M.}~\bibnamefont{Dean}},
  \bibinfo{journal}{Journal of Magnetism and Magnetic Materials}
  \textbf{\bibinfo{volume}{376}}, \bibinfo{pages}{3 } (\bibinfo{year}{2015}),
  ISSN \bibinfo{issn}{0304-8853}.

\bibitem[{\citenamefont{Braicovich et~al.}(2009)\citenamefont{Braicovich,
  Ament, Bisogni, Forte, Aruta, Balestrino, Brookes, De~Luca, Medaglia,
  Granozio et~al.}}]{Braicovich09}
\bibinfo{author}{\bibfnamefont{L.}~\bibnamefont{Braicovich}},
  \bibinfo{author}{\bibfnamefont{L.~J.~P.} \bibnamefont{Ament}},
  \bibinfo{author}{\bibfnamefont{V.}~\bibnamefont{Bisogni}},
  \bibinfo{author}{\bibfnamefont{F.}~\bibnamefont{Forte}},
  \bibinfo{author}{\bibfnamefont{C.}~\bibnamefont{Aruta}},
  \bibinfo{author}{\bibfnamefont{G.}~\bibnamefont{Balestrino}},
  \bibinfo{author}{\bibfnamefont{N.~B.} \bibnamefont{Brookes}},
  \bibinfo{author}{\bibfnamefont{G.~M.} \bibnamefont{De~Luca}},
  \bibinfo{author}{\bibfnamefont{P.~G.} \bibnamefont{Medaglia}},
  \bibinfo{author}{\bibfnamefont{F.~M.} \bibnamefont{Granozio}},
  \bibnamefont{et~al.}, \bibinfo{journal}{Phys. Rev. Lett.}
  \textbf{\bibinfo{volume}{102}}, \bibinfo{pages}{167401}
  (\bibinfo{year}{2009}).

\bibitem[{\citenamefont{Hasan et~al.}(2000)\citenamefont{Hasan, Isaacs, Shen,
  Miller, Tsutsui, Tohyama, and Maekawa}}]{Hasan00}
\bibinfo{author}{\bibfnamefont{M.~Z.} \bibnamefont{Hasan}},
  \bibinfo{author}{\bibfnamefont{E.~D.} \bibnamefont{Isaacs}},
  \bibinfo{author}{\bibfnamefont{Z.-X.} \bibnamefont{Shen}},
  \bibinfo{author}{\bibfnamefont{L.~L.} \bibnamefont{Miller}},
  \bibinfo{author}{\bibfnamefont{K.}~\bibnamefont{Tsutsui}},
  \bibinfo{author}{\bibfnamefont{T.}~\bibnamefont{Tohyama}}, \bibnamefont{and}
  \bibinfo{author}{\bibfnamefont{S.}~\bibnamefont{Maekawa}},
  \bibinfo{journal}{Science} \textbf{\bibinfo{volume}{288}},
  \bibinfo{pages}{1811} (\bibinfo{year}{2000}).

\bibitem[{\citenamefont{Kim et~al.}(2002)\citenamefont{Kim, Hill, Burns,
  Wakimoto, Birgeneau, Casa, Gog, and Venkataraman}}]{YJKim2002}
\bibinfo{author}{\bibfnamefont{Y.~J.} \bibnamefont{Kim}},
  \bibinfo{author}{\bibfnamefont{J.~P.} \bibnamefont{Hill}},
  \bibinfo{author}{\bibfnamefont{C.~A.} \bibnamefont{Burns}},
  \bibinfo{author}{\bibfnamefont{S.}~\bibnamefont{Wakimoto}},
  \bibinfo{author}{\bibfnamefont{R.~J.} \bibnamefont{Birgeneau}},
  \bibinfo{author}{\bibfnamefont{D.}~\bibnamefont{Casa}},
  \bibinfo{author}{\bibfnamefont{T.}~\bibnamefont{Gog}}, \bibnamefont{and}
  \bibinfo{author}{\bibfnamefont{C.~T.} \bibnamefont{Venkataraman}},
  \bibinfo{journal}{Phys. Rev. Lett.} \textbf{\bibinfo{volume}{89}},
  \bibinfo{pages}{177003} (\bibinfo{year}{2002}).

\bibitem[{\citenamefont{Jarrige et~al.}(2010)\citenamefont{Jarrige, Ishii,
  Yoshida, Fukuda, Ikeuchi, Ishikado, Hiraoka, Tsuei, Kito, Iyo
  et~al.}}]{Jarrige_XES}
\bibinfo{author}{\bibfnamefont{I.}~\bibnamefont{Jarrige}},
  \bibinfo{author}{\bibfnamefont{K.}~\bibnamefont{Ishii}},
  \bibinfo{author}{\bibfnamefont{M.}~\bibnamefont{Yoshida}},
  \bibinfo{author}{\bibfnamefont{T.}~\bibnamefont{Fukuda}},
  \bibinfo{author}{\bibfnamefont{K.}~\bibnamefont{Ikeuchi}},
  \bibinfo{author}{\bibfnamefont{M.}~\bibnamefont{Ishikado}},
  \bibinfo{author}{\bibfnamefont{N.}~\bibnamefont{Hiraoka}},
  \bibinfo{author}{\bibfnamefont{K.~D.} \bibnamefont{Tsuei}},
  \bibinfo{author}{\bibfnamefont{H.}~\bibnamefont{Kito}},
  \bibinfo{author}{\bibfnamefont{A.}~\bibnamefont{Iyo}}, \bibnamefont{et~al.},
  \bibinfo{journal}{Physica C} \textbf{\bibinfo{volume}{470}},
  \bibinfo{pages}{S377} (\bibinfo{year}{2010}).

\bibitem[{\citenamefont{Jarrige et~al.}(2012)\citenamefont{Jarrige, Nomura,
  Ishii, Gretarsson, Kim, Kim, Upton, Casa, Gog, Ishikado
  et~al.}}]{Jarrige_RIXS}
\bibinfo{author}{\bibfnamefont{I.}~\bibnamefont{Jarrige}},
  \bibinfo{author}{\bibfnamefont{T.}~\bibnamefont{Nomura}},
  \bibinfo{author}{\bibfnamefont{K.}~\bibnamefont{Ishii}},
  \bibinfo{author}{\bibfnamefont{H.}~\bibnamefont{Gretarsson}},
  \bibinfo{author}{\bibfnamefont{Y.-J.} \bibnamefont{Kim}},
  \bibinfo{author}{\bibfnamefont{J.}~\bibnamefont{Kim}},
  \bibinfo{author}{\bibfnamefont{M.}~\bibnamefont{Upton}},
  \bibinfo{author}{\bibfnamefont{D.}~\bibnamefont{Casa}},
  \bibinfo{author}{\bibfnamefont{T.}~\bibnamefont{Gog}},
  \bibinfo{author}{\bibfnamefont{M.}~\bibnamefont{Ishikado}},
  \bibnamefont{et~al.}, \bibinfo{journal}{Phys. Rev. B}
  \textbf{\bibinfo{volume}{86}}, \bibinfo{pages}{115104}
  (\bibinfo{year}{2012}).

\bibitem[{\citenamefont{Yang et~al.}(2009)\citenamefont{Yang, Sorini, Chen,
  Moritz, Lee, Vernay, Olalde-Velasco, Denlinger, Delley, Chu
  et~al.}}]{Yang2009}
\bibinfo{author}{\bibfnamefont{W.}~\bibnamefont{Yang}},
  \bibinfo{author}{\bibfnamefont{A.}~\bibnamefont{Sorini}},
  \bibinfo{author}{\bibfnamefont{C.-C.} \bibnamefont{Chen}},
  \bibinfo{author}{\bibfnamefont{B.}~\bibnamefont{Moritz}},
  \bibinfo{author}{\bibfnamefont{W.-S.} \bibnamefont{Lee}},
  \bibinfo{author}{\bibfnamefont{F.}~\bibnamefont{Vernay}},
  \bibinfo{author}{\bibfnamefont{P.}~\bibnamefont{Olalde-Velasco}},
  \bibinfo{author}{\bibfnamefont{J.}~\bibnamefont{Denlinger}},
  \bibinfo{author}{\bibfnamefont{B.}~\bibnamefont{Delley}},
  \bibinfo{author}{\bibfnamefont{J.-H.} \bibnamefont{Chu}},
  \bibnamefont{et~al.}, \bibinfo{journal}{Phys. Rev. B}
  \textbf{\bibinfo{volume}{80}}, \bibinfo{pages}{014508}
  (\bibinfo{year}{2009}).

\bibitem[{\citenamefont{Zhou et~al.}(2013)\citenamefont{Zhou, Huang, Monney,
  Dai, Strocov, Wang, Chen, Zhang, Dai, Patthey et~al.}}]{FeL3_RIXS}
\bibinfo{author}{\bibfnamefont{K.-J.} \bibnamefont{Zhou}},
  \bibinfo{author}{\bibfnamefont{Y.-B.} \bibnamefont{Huang}},
  \bibinfo{author}{\bibfnamefont{C.}~\bibnamefont{Monney}},
  \bibinfo{author}{\bibfnamefont{X.}~\bibnamefont{Dai}},
  \bibinfo{author}{\bibfnamefont{V.~N.} \bibnamefont{Strocov}},
  \bibinfo{author}{\bibfnamefont{N.-L.} \bibnamefont{Wang}},
  \bibinfo{author}{\bibfnamefont{Z.-G.} \bibnamefont{Chen}},
  \bibinfo{author}{\bibfnamefont{C.}~\bibnamefont{Zhang}},
  \bibinfo{author}{\bibfnamefont{P.}~\bibnamefont{Dai}},
  \bibinfo{author}{\bibfnamefont{L.}~\bibnamefont{Patthey}},
  \bibnamefont{et~al.}, \bibinfo{journal}{Nat. Commun.}
  \textbf{\bibinfo{volume}{4}}, \bibinfo{pages}{1470} (\bibinfo{year}{2013}).

\bibitem[{\citenamefont{Tacon et~al.}(2011)\citenamefont{Tacon, Ghiringhelli,
  Chaloupka, Sala, Hinkov, Haverkort, Minola, Bakr, Zhou, Blanco-Canosa
  et~al.}}]{Keimer11}
\bibinfo{author}{\bibfnamefont{M.~L.} \bibnamefont{Tacon}},
  \bibinfo{author}{\bibfnamefont{G.}~\bibnamefont{Ghiringhelli}},
  \bibinfo{author}{\bibfnamefont{J.}~\bibnamefont{Chaloupka}},
  \bibinfo{author}{\bibfnamefont{M.~M.} \bibnamefont{Sala}},
  \bibinfo{author}{\bibfnamefont{V.}~\bibnamefont{Hinkov}},
  \bibinfo{author}{\bibfnamefont{M.~W.} \bibnamefont{Haverkort}},
  \bibinfo{author}{\bibfnamefont{M.}~\bibnamefont{Minola}},
  \bibinfo{author}{\bibfnamefont{M.}~\bibnamefont{Bakr}},
  \bibinfo{author}{\bibfnamefont{K.~J.} \bibnamefont{Zhou}},
  \bibinfo{author}{\bibfnamefont{S.}~\bibnamefont{Blanco-Canosa}},
  \bibnamefont{et~al.}, \bibinfo{journal}{Nat. Phys.}
  \textbf{\bibinfo{volume}{7}} (\bibinfo{year}{2011}).

\bibitem[{\citenamefont{Guo et~al.}(2010)\citenamefont{Guo, Jin, Wang, Wang,
  Zhu, Zhou, He, and Chen}}]{Guo_KFe2Se2:2010}
\bibinfo{author}{\bibfnamefont{J.}~\bibnamefont{Guo}},
  \bibinfo{author}{\bibfnamefont{S.}~\bibnamefont{Jin}},
  \bibinfo{author}{\bibfnamefont{G.}~\bibnamefont{Wang}},
  \bibinfo{author}{\bibfnamefont{S.}~\bibnamefont{Wang}},
  \bibinfo{author}{\bibfnamefont{K.}~\bibnamefont{Zhu}},
  \bibinfo{author}{\bibfnamefont{T.}~\bibnamefont{Zhou}},
  \bibinfo{author}{\bibfnamefont{M.}~\bibnamefont{He}}, \bibnamefont{and}
  \bibinfo{author}{\bibfnamefont{X.}~\bibnamefont{Chen}},
  \bibinfo{journal}{Phys. Rev. B} \textbf{\bibinfo{volume}{82}},
  \bibinfo{pages}{180520} (\bibinfo{year}{2010}).

\bibitem[{\citenamefont{Wang et~al.}(2011)\citenamefont{Wang, Ying, Yan, Liu,
  Luo, Li, Wang, Zhang, Ye, Cheng et~al.}}]{Wang_RbFe2Se2:2011}
\bibinfo{author}{\bibfnamefont{A.~F.} \bibnamefont{Wang}},
  \bibinfo{author}{\bibfnamefont{J.~J.} \bibnamefont{Ying}},
  \bibinfo{author}{\bibfnamefont{Y.~J.} \bibnamefont{Yan}},
  \bibinfo{author}{\bibfnamefont{R.~H.} \bibnamefont{Liu}},
  \bibinfo{author}{\bibfnamefont{X.~G.} \bibnamefont{Luo}},
  \bibinfo{author}{\bibfnamefont{Z.~Y.} \bibnamefont{Li}},
  \bibinfo{author}{\bibfnamefont{X.~F.} \bibnamefont{Wang}},
  \bibinfo{author}{\bibfnamefont{M.}~\bibnamefont{Zhang}},
  \bibinfo{author}{\bibfnamefont{G.~J.} \bibnamefont{Ye}},
  \bibinfo{author}{\bibfnamefont{P.}~\bibnamefont{Cheng}},
  \bibnamefont{et~al.}, \bibinfo{journal}{Phys. Rev. B}
  \textbf{\bibinfo{volume}{83}}, \bibinfo{pages}{060512}
  (\bibinfo{year}{2011}).

\bibitem[{\citenamefont{Krzton-Maziopa
  et~al.}(2011)\citenamefont{Krzton-Maziopa, Shermadini, Pomjakushina,
  Pomjakushin, Bendele, Amato, Khasanov, Luetkens, and
  Conder}}]{Krzton_CsFe2Se2:2011}
\bibinfo{author}{\bibfnamefont{A.}~\bibnamefont{Krzton-Maziopa}},
  \bibinfo{author}{\bibfnamefont{Z.}~\bibnamefont{Shermadini}},
  \bibinfo{author}{\bibfnamefont{E.}~\bibnamefont{Pomjakushina}},
  \bibinfo{author}{\bibfnamefont{V.}~\bibnamefont{Pomjakushin}},
  \bibinfo{author}{\bibfnamefont{M.}~\bibnamefont{Bendele}},
  \bibinfo{author}{\bibfnamefont{A.}~\bibnamefont{Amato}},
  \bibinfo{author}{\bibfnamefont{R.}~\bibnamefont{Khasanov}},
  \bibinfo{author}{\bibfnamefont{H.}~\bibnamefont{Luetkens}}, \bibnamefont{and}
  \bibinfo{author}{\bibfnamefont{K.}~\bibnamefont{Conder}},
  \bibinfo{journal}{Journal of Physics: Condensed Matter}
  \textbf{\bibinfo{volume}{23}}, \bibinfo{pages}{052203}
  (\bibinfo{year}{2011}).

\bibitem[{\citenamefont{Dagotto}(2013)}]{Dagotto_RMP:2013}
\bibinfo{author}{\bibfnamefont{E.}~\bibnamefont{Dagotto}},
  \bibinfo{journal}{Rev. Mod. Phys.} \textbf{\bibinfo{volume}{85}},
  \bibinfo{pages}{849} (\bibinfo{year}{2013}).

\bibitem[{\citenamefont{Johnston}(2010)}]{Johnston2010}
\bibinfo{author}{\bibfnamefont{D.~C.} \bibnamefont{Johnston}},
  \bibinfo{journal}{Advances in Physics} \textbf{\bibinfo{volume}{59}},
  \bibinfo{pages}{803} (\bibinfo{year}{2010}).

\bibitem[{\citenamefont{Kuroki et~al.}(2008)\citenamefont{Kuroki, Onari, Arita,
  Usui, Tanaka, Kontani, and Aoki}}]{Kuroki_PRL:2008}
\bibinfo{author}{\bibfnamefont{K.}~\bibnamefont{Kuroki}},
  \bibinfo{author}{\bibfnamefont{S.}~\bibnamefont{Onari}},
  \bibinfo{author}{\bibfnamefont{R.}~\bibnamefont{Arita}},
  \bibinfo{author}{\bibfnamefont{H.}~\bibnamefont{Usui}},
  \bibinfo{author}{\bibfnamefont{Y.}~\bibnamefont{Tanaka}},
  \bibinfo{author}{\bibfnamefont{H.}~\bibnamefont{Kontani}}, \bibnamefont{and}
  \bibinfo{author}{\bibfnamefont{H.}~\bibnamefont{Aoki}},
  \bibinfo{journal}{Phys. Rev. Lett.} \textbf{\bibinfo{volume}{101}},
  \bibinfo{pages}{087004} (\bibinfo{year}{2008}).

\bibitem[{\citenamefont{Zhang et~al.}(2011)\citenamefont{Zhang, Yang, Xu, Ye,
  Chen, He, Xu, Jiang, Xie, Ying et~al.}}]{Feng_NatMat:2011}
\bibinfo{author}{\bibfnamefont{Y.}~\bibnamefont{Zhang}},
  \bibinfo{author}{\bibfnamefont{L.~X.} \bibnamefont{Yang}},
  \bibinfo{author}{\bibfnamefont{M.}~\bibnamefont{Xu}},
  \bibinfo{author}{\bibfnamefont{Z.~R.} \bibnamefont{Ye}},
  \bibinfo{author}{\bibfnamefont{F.}~\bibnamefont{Chen}},
  \bibinfo{author}{\bibfnamefont{C.}~\bibnamefont{He}},
  \bibinfo{author}{\bibfnamefont{H.~C.} \bibnamefont{Xu}},
  \bibinfo{author}{\bibfnamefont{J.}~\bibnamefont{Jiang}},
  \bibinfo{author}{\bibfnamefont{B.~P.} \bibnamefont{Xie}},
  \bibinfo{author}{\bibfnamefont{J.~J.} \bibnamefont{Ying}},
  \bibnamefont{et~al.}, \bibinfo{journal}{Nat. Mater.}
  \textbf{\bibinfo{volume}{10}}, \bibinfo{pages}{273} (\bibinfo{year}{2011}).

\bibitem[{\citenamefont{Qian et~al.}(2011{\natexlab{a}})\citenamefont{Qian,
  Wang, Jin, Zhang, Richard, Xu, Dai, Fang, Guo, Chen et~al.}}]{Ding_PRL:2011}
\bibinfo{author}{\bibfnamefont{T.}~\bibnamefont{Qian}},
  \bibinfo{author}{\bibfnamefont{X.-P.} \bibnamefont{Wang}},
  \bibinfo{author}{\bibfnamefont{W.-C.} \bibnamefont{Jin}},
  \bibinfo{author}{\bibfnamefont{P.}~\bibnamefont{Zhang}},
  \bibinfo{author}{\bibfnamefont{P.}~\bibnamefont{Richard}},
  \bibinfo{author}{\bibfnamefont{G.}~\bibnamefont{Xu}},
  \bibinfo{author}{\bibfnamefont{X.}~\bibnamefont{Dai}},
  \bibinfo{author}{\bibfnamefont{Z.}~\bibnamefont{Fang}},
  \bibinfo{author}{\bibfnamefont{J.-G.} \bibnamefont{Guo}},
  \bibinfo{author}{\bibfnamefont{X.-L.} \bibnamefont{Chen}},
  \bibnamefont{et~al.}, \bibinfo{journal}{Phys. Rev. Lett.}
  \textbf{\bibinfo{volume}{106}}, \bibinfo{pages}{187001}
  (\bibinfo{year}{2011}{\natexlab{a}}).

\bibitem[{\citenamefont{Mou et~al.}(2011)\citenamefont{Mou, Liu, Jia, He, Peng,
  Zhao, Yu, Liu, He, Dong et~al.}}]{Zhou_PRL:2011}
\bibinfo{author}{\bibfnamefont{D.}~\bibnamefont{Mou}},
  \bibinfo{author}{\bibfnamefont{S.}~\bibnamefont{Liu}},
  \bibinfo{author}{\bibfnamefont{X.}~\bibnamefont{Jia}},
  \bibinfo{author}{\bibfnamefont{J.}~\bibnamefont{He}},
  \bibinfo{author}{\bibfnamefont{Y.}~\bibnamefont{Peng}},
  \bibinfo{author}{\bibfnamefont{L.}~\bibnamefont{Zhao}},
  \bibinfo{author}{\bibfnamefont{L.}~\bibnamefont{Yu}},
  \bibinfo{author}{\bibfnamefont{G.}~\bibnamefont{Liu}},
  \bibinfo{author}{\bibfnamefont{S.}~\bibnamefont{He}},
  \bibinfo{author}{\bibfnamefont{X.}~\bibnamefont{Dong}}, \bibnamefont{et~al.},
  \bibinfo{journal}{Phys. Rev. Lett.} \textbf{\bibinfo{volume}{106}},
  \bibinfo{pages}{107001} (\bibinfo{year}{2011}).

\bibitem[{\citenamefont{Mazin et~al.}(2008)\citenamefont{Mazin, Singh,
  Johannes, and Du}}]{Mazin_PRL:2008}
\bibinfo{author}{\bibfnamefont{I.~I.} \bibnamefont{Mazin}},
  \bibinfo{author}{\bibfnamefont{D.~J.} \bibnamefont{Singh}},
  \bibinfo{author}{\bibfnamefont{M.~D.} \bibnamefont{Johannes}},
  \bibnamefont{and} \bibinfo{author}{\bibfnamefont{M.~H.} \bibnamefont{Du}},
  \bibinfo{journal}{Phys. Rev. Lett.} \textbf{\bibinfo{volume}{101}},
  \bibinfo{pages}{057003} (\bibinfo{year}{2008}).

\bibitem[{\citenamefont{Chubukov et~al.}(2008)\citenamefont{Chubukov, Efremov,
  and Eremin}}]{Chubukov_PRB:2008}
\bibinfo{author}{\bibfnamefont{A.~V.} \bibnamefont{Chubukov}},
  \bibinfo{author}{\bibfnamefont{D.~V.} \bibnamefont{Efremov}},
  \bibnamefont{and} \bibinfo{author}{\bibfnamefont{I.}~\bibnamefont{Eremin}},
  \bibinfo{journal}{Phys. Rev. B} \textbf{\bibinfo{volume}{78}},
  \bibinfo{pages}{134512} (\bibinfo{year}{2008}).

\bibitem[{\citenamefont{Wang et~al.}(2009)\citenamefont{Wang, Zhai, Ran,
  Vishwanath, and Lee}}]{FaWang_PRL:2009}
\bibinfo{author}{\bibfnamefont{F.}~\bibnamefont{Wang}},
  \bibinfo{author}{\bibfnamefont{H.}~\bibnamefont{Zhai}},
  \bibinfo{author}{\bibfnamefont{Y.}~\bibnamefont{Ran}},
  \bibinfo{author}{\bibfnamefont{A.}~\bibnamefont{Vishwanath}},
  \bibnamefont{and} \bibinfo{author}{\bibfnamefont{D.-H.} \bibnamefont{Lee}},
  \bibinfo{journal}{Phys. Rev. Lett.} \textbf{\bibinfo{volume}{102}},
  \bibinfo{pages}{047005} (\bibinfo{year}{2009}).

\bibitem[{\citenamefont{Fang et~al.}(2011)\citenamefont{Fang, Wang, Dong, Li,
  Feng, Chen, and Yuan}}]{FangEPL_PhaseDiagram:2011}
\bibinfo{author}{\bibfnamefont{M.-H.} \bibnamefont{Fang}},
  \bibinfo{author}{\bibfnamefont{H.-D.} \bibnamefont{Wang}},
  \bibinfo{author}{\bibfnamefont{C.-H.} \bibnamefont{Dong}},
  \bibinfo{author}{\bibfnamefont{Z.-J.} \bibnamefont{Li}},
  \bibinfo{author}{\bibfnamefont{C.-M.} \bibnamefont{Feng}},
  \bibinfo{author}{\bibfnamefont{J.}~\bibnamefont{Chen}}, \bibnamefont{and}
  \bibinfo{author}{\bibfnamefont{H.~Q.} \bibnamefont{Yuan}},
  \bibinfo{journal}{EPL (Europhysics Letters)} \textbf{\bibinfo{volume}{94}},
  \bibinfo{pages}{27009} (\bibinfo{year}{2011}).

\bibitem[{\citenamefont{Bao et~al.}(2011)\citenamefont{Bao, Huang, Chen, Green,
  Wang, He, and Qiu}}]{WeiBao_KFe2Se2_Neutron:2011}
\bibinfo{author}{\bibfnamefont{W.}~\bibnamefont{Bao}},
  \bibinfo{author}{\bibfnamefont{Q.-Z.} \bibnamefont{Huang}},
  \bibinfo{author}{\bibfnamefont{G.-F.} \bibnamefont{Chen}},
  \bibinfo{author}{\bibfnamefont{M.~A.} \bibnamefont{Green}},
  \bibinfo{author}{\bibfnamefont{D.-M.} \bibnamefont{Wang}},
  \bibinfo{author}{\bibfnamefont{J.-B.} \bibnamefont{He}}, \bibnamefont{and}
  \bibinfo{author}{\bibfnamefont{Y.-M.} \bibnamefont{Qiu}},
  \bibinfo{journal}{Chinese Physics Letters} \textbf{\bibinfo{volume}{28}},
  \bibinfo{pages}{086104} (\bibinfo{year}{2011}).

\bibitem[{\citenamefont{Chen et~al.}(2011{\natexlab{a}})\citenamefont{Chen, Xu,
  Ge, Zhang, Ye, Yang, Jiang, Xie, Che, Zhang et~al.}}]{Chen2011}
\bibinfo{author}{\bibfnamefont{F.}~\bibnamefont{Chen}},
  \bibinfo{author}{\bibfnamefont{M.}~\bibnamefont{Xu}},
  \bibinfo{author}{\bibfnamefont{Q.~Q.} \bibnamefont{Ge}},
  \bibinfo{author}{\bibfnamefont{Y.}~\bibnamefont{Zhang}},
  \bibinfo{author}{\bibfnamefont{Z.~R.} \bibnamefont{Ye}},
  \bibinfo{author}{\bibfnamefont{L.~X.} \bibnamefont{Yang}},
  \bibinfo{author}{\bibfnamefont{J.}~\bibnamefont{Jiang}},
  \bibinfo{author}{\bibfnamefont{B.~P.} \bibnamefont{Xie}},
  \bibinfo{author}{\bibfnamefont{R.~C.} \bibnamefont{Che}},
  \bibinfo{author}{\bibfnamefont{M.}~\bibnamefont{Zhang}},
  \bibnamefont{et~al.}, \bibinfo{journal}{Phys. Rev. X}
  \textbf{\bibinfo{volume}{1}}, \bibinfo{pages}{021020}
  (\bibinfo{year}{2011}{\natexlab{a}}).

\bibitem[{\citenamefont{Ricci et~al.}(2011)\citenamefont{Ricci, Poccia, Campi,
  Joseph, Arrighetti, Barba, Reynolds, Burghammer, Takeya, Mizuguchi
  et~al.}}]{Bianconi:2011}
\bibinfo{author}{\bibfnamefont{A.}~\bibnamefont{Ricci}},
  \bibinfo{author}{\bibfnamefont{N.}~\bibnamefont{Poccia}},
  \bibinfo{author}{\bibfnamefont{G.}~\bibnamefont{Campi}},
  \bibinfo{author}{\bibfnamefont{B.}~\bibnamefont{Joseph}},
  \bibinfo{author}{\bibfnamefont{G.}~\bibnamefont{Arrighetti}},
  \bibinfo{author}{\bibfnamefont{L.}~\bibnamefont{Barba}},
  \bibinfo{author}{\bibfnamefont{M.}~\bibnamefont{Reynolds}},
  \bibinfo{author}{\bibfnamefont{M.}~\bibnamefont{Burghammer}},
  \bibinfo{author}{\bibfnamefont{H.}~\bibnamefont{Takeya}},
  \bibinfo{author}{\bibfnamefont{Y.}~\bibnamefont{Mizuguchi}},
  \bibnamefont{et~al.}, \bibinfo{journal}{Phys. Rev. B}
  \textbf{\bibinfo{volume}{84}}, \bibinfo{pages}{060511}
  (\bibinfo{year}{2011}).

\bibitem[{\citenamefont{Ksenofontov et~al.}(2011)\citenamefont{Ksenofontov,
  Wortmann, Medvedev, Tsurkan, Deisenhofer, Loidl, and
  Felser}}]{Deisenhofer:2011}
\bibinfo{author}{\bibfnamefont{V.}~\bibnamefont{Ksenofontov}},
  \bibinfo{author}{\bibfnamefont{G.}~\bibnamefont{Wortmann}},
  \bibinfo{author}{\bibfnamefont{S.~A.} \bibnamefont{Medvedev}},
  \bibinfo{author}{\bibfnamefont{V.}~\bibnamefont{Tsurkan}},
  \bibinfo{author}{\bibfnamefont{J.}~\bibnamefont{Deisenhofer}},
  \bibinfo{author}{\bibfnamefont{A.}~\bibnamefont{Loidl}}, \bibnamefont{and}
  \bibinfo{author}{\bibfnamefont{C.}~\bibnamefont{Felser}},
  \bibinfo{journal}{Phys. Rev. B} \textbf{\bibinfo{volume}{84}},
  \bibinfo{pages}{180508} (\bibinfo{year}{2011}).

\bibitem[{\citenamefont{Shermadini et~al.}(2011)\citenamefont{Shermadini,
  Krzton-Maziopa, Bendele, Khasanov, Luetkens, Conder, Pomjakushina, Weyeneth,
  Pomjakushin, Bossen et~al.}}]{Amato:2011}
\bibinfo{author}{\bibfnamefont{Z.}~\bibnamefont{Shermadini}},
  \bibinfo{author}{\bibfnamefont{A.}~\bibnamefont{Krzton-Maziopa}},
  \bibinfo{author}{\bibfnamefont{M.}~\bibnamefont{Bendele}},
  \bibinfo{author}{\bibfnamefont{R.}~\bibnamefont{Khasanov}},
  \bibinfo{author}{\bibfnamefont{H.}~\bibnamefont{Luetkens}},
  \bibinfo{author}{\bibfnamefont{K.}~\bibnamefont{Conder}},
  \bibinfo{author}{\bibfnamefont{E.}~\bibnamefont{Pomjakushina}},
  \bibinfo{author}{\bibfnamefont{S.}~\bibnamefont{Weyeneth}},
  \bibinfo{author}{\bibfnamefont{V.}~\bibnamefont{Pomjakushin}},
  \bibinfo{author}{\bibfnamefont{O.}~\bibnamefont{Bossen}},
  \bibnamefont{et~al.}, \bibinfo{journal}{Phys. Rev. Lett.}
  \textbf{\bibinfo{volume}{106}}, \bibinfo{pages}{117602}
  (\bibinfo{year}{2011}).

\bibitem[{\citenamefont{Li et~al.}(2012)\citenamefont{Li, Ding, Li, Deng,
  Chang, He, Ji, Wang, Ma, Hu et~al.}}]{WeiPRL_STM:2012}
\bibinfo{author}{\bibfnamefont{W.}~\bibnamefont{Li}},
  \bibinfo{author}{\bibfnamefont{H.}~\bibnamefont{Ding}},
  \bibinfo{author}{\bibfnamefont{Z.}~\bibnamefont{Li}},
  \bibinfo{author}{\bibfnamefont{P.}~\bibnamefont{Deng}},
  \bibinfo{author}{\bibfnamefont{K.}~\bibnamefont{Chang}},
  \bibinfo{author}{\bibfnamefont{K.}~\bibnamefont{He}},
  \bibinfo{author}{\bibfnamefont{S.}~\bibnamefont{Ji}},
  \bibinfo{author}{\bibfnamefont{L.}~\bibnamefont{Wang}},
  \bibinfo{author}{\bibfnamefont{X.}~\bibnamefont{Ma}},
  \bibinfo{author}{\bibfnamefont{J.-P.} \bibnamefont{Hu}},
  \bibnamefont{et~al.}, \bibinfo{journal}{Phys. Rev. Lett.}
  \textbf{\bibinfo{volume}{109}}, \bibinfo{pages}{057003}
  (\bibinfo{year}{2012}).

\bibitem[{\citenamefont{Chen et~al.}(2011{\natexlab{b}})\citenamefont{Chen,
  Yuan, Dong, Xu, Shi, Zheng, Luo, Guo, Chen, and Wang}}]{KFe2Se2_Optics:2011}
\bibinfo{author}{\bibfnamefont{Z.~G.} \bibnamefont{Chen}},
  \bibinfo{author}{\bibfnamefont{R.~H.} \bibnamefont{Yuan}},
  \bibinfo{author}{\bibfnamefont{T.}~\bibnamefont{Dong}},
  \bibinfo{author}{\bibfnamefont{G.}~\bibnamefont{Xu}},
  \bibinfo{author}{\bibfnamefont{Y.~G.} \bibnamefont{Shi}},
  \bibinfo{author}{\bibfnamefont{P.}~\bibnamefont{Zheng}},
  \bibinfo{author}{\bibfnamefont{J.~L.} \bibnamefont{Luo}},
  \bibinfo{author}{\bibfnamefont{J.~G.} \bibnamefont{Guo}},
  \bibinfo{author}{\bibfnamefont{X.~L.} \bibnamefont{Chen}}, \bibnamefont{and}
  \bibinfo{author}{\bibfnamefont{N.~L.} \bibnamefont{Wang}},
  \bibinfo{journal}{Phys. Rev. B} \textbf{\bibinfo{volume}{83}},
  \bibinfo{pages}{220507} (\bibinfo{year}{2011}{\natexlab{b}}).

\bibitem[{\citenamefont{Ye et~al.}(2011)\citenamefont{Ye, Chi, Bao, Wang, Ying,
  Chen, Wang, Dong, and Fang}}]{Ye_PRL:2011}
\bibinfo{author}{\bibfnamefont{F.}~\bibnamefont{Ye}},
  \bibinfo{author}{\bibfnamefont{S.}~\bibnamefont{Chi}},
  \bibinfo{author}{\bibfnamefont{W.}~\bibnamefont{Bao}},
  \bibinfo{author}{\bibfnamefont{X.~F.} \bibnamefont{Wang}},
  \bibinfo{author}{\bibfnamefont{J.~J.} \bibnamefont{Ying}},
  \bibinfo{author}{\bibfnamefont{X.~H.} \bibnamefont{Chen}},
  \bibinfo{author}{\bibfnamefont{H.~D.} \bibnamefont{Wang}},
  \bibinfo{author}{\bibfnamefont{C.~H.} \bibnamefont{Dong}}, \bibnamefont{and}
  \bibinfo{author}{\bibfnamefont{M.}~\bibnamefont{Fang}},
  \bibinfo{journal}{Phys. Rev. Lett.} \textbf{\bibinfo{volume}{107}},
  \bibinfo{pages}{137003} (\bibinfo{year}{2011}).

\bibitem[{\citenamefont{Joseph et~al.}(2010)\citenamefont{Joseph, Iadecola,
  Simonelli, Mizuguchi, Takano, Mizokawa, and Saini}}]{Saini_XAS}
\bibinfo{author}{\bibfnamefont{B.}~\bibnamefont{Joseph}},
  \bibinfo{author}{\bibfnamefont{A.}~\bibnamefont{Iadecola}},
  \bibinfo{author}{\bibfnamefont{L.}~\bibnamefont{Simonelli}},
  \bibinfo{author}{\bibfnamefont{Y.}~\bibnamefont{Mizuguchi}},
  \bibinfo{author}{\bibfnamefont{Y.}~\bibnamefont{Takano}},
  \bibinfo{author}{\bibfnamefont{T.}~\bibnamefont{Mizokawa}}, \bibnamefont{and}
  \bibinfo{author}{\bibfnamefont{N.~L.} \bibnamefont{Saini}},
  \bibinfo{journal}{Journal of Physics: Condensed Matter}
  \textbf{\bibinfo{volume}{22}}, \bibinfo{pages}{485702}
  (\bibinfo{year}{2010}).

\bibitem[{\citenamefont{Ellis et~al.}(2010)\citenamefont{Ellis, Kim, Hill,
  Wakimoto, Birgeneau, Shvyd'ko, Casa, Gog, Ishii, Ikeuchi et~al.}}]{Ellis2010}
\bibinfo{author}{\bibfnamefont{D.}~\bibnamefont{Ellis}},
  \bibinfo{author}{\bibfnamefont{J.}~\bibnamefont{Kim}},
  \bibinfo{author}{\bibfnamefont{J.}~\bibnamefont{Hill}},
  \bibinfo{author}{\bibfnamefont{S.}~\bibnamefont{Wakimoto}},
  \bibinfo{author}{\bibfnamefont{R.}~\bibnamefont{Birgeneau}},
  \bibinfo{author}{\bibfnamefont{Y.}~\bibnamefont{Shvyd'ko}},
  \bibinfo{author}{\bibfnamefont{D.}~\bibnamefont{Casa}},
  \bibinfo{author}{\bibfnamefont{T.}~\bibnamefont{Gog}},
  \bibinfo{author}{\bibfnamefont{K.}~\bibnamefont{Ishii}},
  \bibinfo{author}{\bibfnamefont{K.}~\bibnamefont{Ikeuchi}},
  \bibnamefont{et~al.}, \bibinfo{journal}{Phys. Rev. B}
  \textbf{\bibinfo{volume}{81}}, \bibinfo{pages}{085124}
  (\bibinfo{year}{2010}).

\bibitem[{Cod()}]{Code}
\bibinfo{note}{P. Blaha, K. Schwarz, G.Madsen, D. Kvasnicka and J. Luitz:
  WIEN2k, An Augmented PlaneWave Plus Local Orbitals Program for
  Calculating Crystal Properties (ISBN 3-9501031-1-2).}

\bibitem[{\citenamefont{Mostofi et~al.}(2008)\citenamefont{Mostofi, Yates, Lee,
  Souza, Vanderbilt, and Marzari}}]{Mostofi2008}
\bibinfo{author}{\bibfnamefont{A.~A.} \bibnamefont{Mostofi}},
  \bibinfo{author}{\bibfnamefont{J.~R.} \bibnamefont{Yates}},
  \bibinfo{author}{\bibfnamefont{Y.-S.} \bibnamefont{Lee}},
  \bibinfo{author}{\bibfnamefont{I.}~\bibnamefont{Souza}},
  \bibinfo{author}{\bibfnamefont{D.}~\bibnamefont{Vanderbilt}},
  \bibnamefont{and} \bibinfo{author}{\bibfnamefont{N.}~\bibnamefont{Marzari}},
  \bibinfo{journal}{Computer Physics Communications}
  \textbf{\bibinfo{volume}{178}}, \bibinfo{pages}{685 } (\bibinfo{year}{2008}).

\bibitem[{\citenamefont{Nomura and Igarashi}(2005)}]{Nomura2005}
\bibinfo{author}{\bibfnamefont{T.}~\bibnamefont{Nomura}} \bibnamefont{and}
  \bibinfo{author}{\bibfnamefont{J.-i.} \bibnamefont{Igarashi}},
  \bibinfo{journal}{Phys. Rev. B} \textbf{\bibinfo{volume}{71}},
  \bibinfo{pages}{035110} (\bibinfo{year}{2005}).

\bibitem[{\citenamefont{Takahashi et~al.}(2007)\citenamefont{Takahashi,
  Igarashi, and Nomura}}]{Takahashi2007}
\bibinfo{author}{\bibfnamefont{M.}~\bibnamefont{Takahashi}},
  \bibinfo{author}{\bibfnamefont{J.-I.} \bibnamefont{Igarashi}},
  \bibnamefont{and} \bibinfo{author}{\bibfnamefont{T.}~\bibnamefont{Nomura}},
  \bibinfo{journal}{Phys. Rev. B} \textbf{\bibinfo{volume}{75}},
  \bibinfo{pages}{235113} (\bibinfo{year}{2007}).

\bibitem[{\citenamefont{Ding et~al.}(2011)\citenamefont{Ding, Nakayama,
  Richard, Souma, Sato, Takahashi, Neupane, Xu, Pan, Fedorov
  et~al.}}]{DingARPES_Ba122:2011}
\bibinfo{author}{\bibfnamefont{H.}~\bibnamefont{Ding}},
  \bibinfo{author}{\bibfnamefont{K.}~\bibnamefont{Nakayama}},
  \bibinfo{author}{\bibfnamefont{P.}~\bibnamefont{Richard}},
  \bibinfo{author}{\bibfnamefont{S.}~\bibnamefont{Souma}},
  \bibinfo{author}{\bibfnamefont{T.}~\bibnamefont{Sato}},
  \bibinfo{author}{\bibfnamefont{T.}~\bibnamefont{Takahashi}},
  \bibinfo{author}{\bibfnamefont{M.}~\bibnamefont{Neupane}},
  \bibinfo{author}{\bibfnamefont{Y.-M.} \bibnamefont{Xu}},
  \bibinfo{author}{\bibfnamefont{Z.-H.} \bibnamefont{Pan}},
  \bibinfo{author}{\bibfnamefont{A.~V.} \bibnamefont{Fedorov}},
  \bibnamefont{et~al.}, \bibinfo{journal}{Journal of Physics: Condensed Matter}
  \textbf{\bibinfo{volume}{23}}, \bibinfo{pages}{135701}
  (\bibinfo{year}{2011}).

\bibitem[{\citenamefont{Qian et~al.}(2011{\natexlab{b}})\citenamefont{Qian,
  Wang, Jin, Zhang, Richard, Xu, Dai, Fang, Guo, Chen
  et~al.}}]{DingARPES_K122:2011}
\bibinfo{author}{\bibfnamefont{T.}~\bibnamefont{Qian}},
  \bibinfo{author}{\bibfnamefont{X.-P.} \bibnamefont{Wang}},
  \bibinfo{author}{\bibfnamefont{W.-C.} \bibnamefont{Jin}},
  \bibinfo{author}{\bibfnamefont{P.}~\bibnamefont{Zhang}},
  \bibinfo{author}{\bibfnamefont{P.}~\bibnamefont{Richard}},
  \bibinfo{author}{\bibfnamefont{G.}~\bibnamefont{Xu}},
  \bibinfo{author}{\bibfnamefont{X.}~\bibnamefont{Dai}},
  \bibinfo{author}{\bibfnamefont{Z.}~\bibnamefont{Fang}},
  \bibinfo{author}{\bibfnamefont{J.-G.} \bibnamefont{Guo}},
  \bibinfo{author}{\bibfnamefont{X.-L.} \bibnamefont{Chen}},
  \bibnamefont{et~al.}, \bibinfo{journal}{Phys. Rev. Lett.}
  \textbf{\bibinfo{volume}{106}}, \bibinfo{pages}{187001}
  (\bibinfo{year}{2011}{\natexlab{b}}).

\bibitem[{\citenamefont{Nomura}(2014)}]{NomuraRIXS:2014}
\bibinfo{author}{\bibfnamefont{T.}~\bibnamefont{Nomura}}, \bibinfo{journal}{J.
  Phys. Soc. Jpn.} \textbf{\bibinfo{volume}{83}}, \bibinfo{pages}{064704}
  (\bibinfo{year}{2014}).

\bibitem[{\citenamefont{Luo et~al.}(2011)\citenamefont{Luo, Nicholson, Riera,
  Yao, Moreo, and Dagotto}}]{Dagotto_PRB:2011}
\bibinfo{author}{\bibfnamefont{Q.}~\bibnamefont{Luo}},
  \bibinfo{author}{\bibfnamefont{A.}~\bibnamefont{Nicholson}},
  \bibinfo{author}{\bibfnamefont{J.}~\bibnamefont{Riera}},
  \bibinfo{author}{\bibfnamefont{D.-X.} \bibnamefont{Yao}},
  \bibinfo{author}{\bibfnamefont{A.}~\bibnamefont{Moreo}}, \bibnamefont{and}
  \bibinfo{author}{\bibfnamefont{E.}~\bibnamefont{Dagotto}},
  \bibinfo{journal}{Phys. Rev. B} \textbf{\bibinfo{volume}{84}},
  \bibinfo{pages}{140506} (\bibinfo{year}{2011}).

\bibitem[{\citenamefont{Chu et~al.}(2010)\citenamefont{Chu, Analytis, De~Greve,
  McMahon, Islam, Yamamoto, and Fisher}}]{Fisher2010}
\bibinfo{author}{\bibfnamefont{J.-H.} \bibnamefont{Chu}},
  \bibinfo{author}{\bibfnamefont{J.~G.} \bibnamefont{Analytis}},
  \bibinfo{author}{\bibfnamefont{K.}~\bibnamefont{De~Greve}},
  \bibinfo{author}{\bibfnamefont{P.~L.} \bibnamefont{McMahon}},
  \bibinfo{author}{\bibfnamefont{Z.}~\bibnamefont{Islam}},
  \bibinfo{author}{\bibfnamefont{Y.}~\bibnamefont{Yamamoto}}, \bibnamefont{and}
  \bibinfo{author}{\bibfnamefont{I.~R.} \bibnamefont{Fisher}},
  \bibinfo{journal}{Science} \textbf{\bibinfo{volume}{329}},
  \bibinfo{pages}{824} (\bibinfo{year}{2010}).

\bibitem[{\citenamefont{Yi et~al.}(2011)\citenamefont{Yi, Lu, Chu, Analytis,
  Sorini, Kemper, Moritz, Mo, Moore, Hashimoto et~al.}}]{Shen2011}
\bibinfo{author}{\bibfnamefont{M.}~\bibnamefont{Yi}},
  \bibinfo{author}{\bibfnamefont{D.}~\bibnamefont{Lu}},
  \bibinfo{author}{\bibfnamefont{J.-H.} \bibnamefont{Chu}},
  \bibinfo{author}{\bibfnamefont{J.~G.} \bibnamefont{Analytis}},
  \bibinfo{author}{\bibfnamefont{A.~P.} \bibnamefont{Sorini}},
  \bibinfo{author}{\bibfnamefont{A.~F.} \bibnamefont{Kemper}},
  \bibinfo{author}{\bibfnamefont{B.}~\bibnamefont{Moritz}},
  \bibinfo{author}{\bibfnamefont{S.-K.} \bibnamefont{Mo}},
  \bibinfo{author}{\bibfnamefont{R.~G.} \bibnamefont{Moore}},
  \bibinfo{author}{\bibfnamefont{M.}~\bibnamefont{Hashimoto}},
  \bibnamefont{et~al.}, \bibinfo{journal}{Proceedings of the National Academy
  of Sciences} \textbf{\bibinfo{volume}{108}}, \bibinfo{pages}{6878}
  (\bibinfo{year}{2011}).

\bibitem[{\citenamefont{Kasahara et~al.}(2012)\citenamefont{Kasahara, Shi,
  Hashimoto, Tonegawa, Mizukami, Shibauchi, Sugimoto, Fukuda, Terashima,
  Nevidomskyy et~al.}}]{Kasahara2012}
\bibinfo{author}{\bibfnamefont{S.}~\bibnamefont{Kasahara}},
  \bibinfo{author}{\bibfnamefont{H.~J.} \bibnamefont{Shi}},
  \bibinfo{author}{\bibfnamefont{K.}~\bibnamefont{Hashimoto}},
  \bibinfo{author}{\bibfnamefont{S.}~\bibnamefont{Tonegawa}},
  \bibinfo{author}{\bibfnamefont{Y.}~\bibnamefont{Mizukami}},
  \bibinfo{author}{\bibfnamefont{T.}~\bibnamefont{Shibauchi}},
  \bibinfo{author}{\bibfnamefont{K.}~\bibnamefont{Sugimoto}},
  \bibinfo{author}{\bibfnamefont{T.}~\bibnamefont{Fukuda}},
  \bibinfo{author}{\bibfnamefont{T.}~\bibnamefont{Terashima}},
  \bibinfo{author}{\bibfnamefont{A.~H.} \bibnamefont{Nevidomskyy}},
  \bibnamefont{et~al.}, \bibinfo{journal}{Nature}
  \textbf{\bibinfo{volume}{486}}, \bibinfo{pages}{382} (\bibinfo{year}{2012}).

\bibitem[{\citenamefont{Wang et~al.}(2014)\citenamefont{Wang, Schmidt, Fischer,
  Tsurkan, Greger, Vollhardt, Loidl, and Deisenhofer}}]{WangNatComm_OSMI:2014}
\bibinfo{author}{\bibfnamefont{Z.}~\bibnamefont{Wang}},
  \bibinfo{author}{\bibfnamefont{M.}~\bibnamefont{Schmidt}},
  \bibinfo{author}{\bibfnamefont{J.}~\bibnamefont{Fischer}},
  \bibinfo{author}{\bibfnamefont{V.}~\bibnamefont{Tsurkan}},
  \bibinfo{author}{\bibfnamefont{M.}~\bibnamefont{Greger}},
  \bibinfo{author}{\bibfnamefont{D.}~\bibnamefont{Vollhardt}},
  \bibinfo{author}{\bibfnamefont{A.}~\bibnamefont{Loidl}}, \bibnamefont{and}
  \bibinfo{author}{\bibfnamefont{J.}~\bibnamefont{Deisenhofer}},
  \bibinfo{journal}{Nat Commun} \textbf{\bibinfo{volume}{5}},
  \bibinfo{pages}{3202} (\bibinfo{year}{2014}).

\bibitem[{\citenamefont{Yi et~al.}(2013)\citenamefont{Yi, Lu, Yu, Riggs, Chu,
  Lv, Liu, Lu, Cui, Hashimoto et~al.}}]{YiPRl_OSMI:2013}
\bibinfo{author}{\bibfnamefont{M.}~\bibnamefont{Yi}},
  \bibinfo{author}{\bibfnamefont{D.~H.} \bibnamefont{Lu}},
  \bibinfo{author}{\bibfnamefont{R.}~\bibnamefont{Yu}},
  \bibinfo{author}{\bibfnamefont{S.~C.} \bibnamefont{Riggs}},
  \bibinfo{author}{\bibfnamefont{J.-H.} \bibnamefont{Chu}},
  \bibinfo{author}{\bibfnamefont{B.}~\bibnamefont{Lv}},
  \bibinfo{author}{\bibfnamefont{Z.~K.} \bibnamefont{Liu}},
  \bibinfo{author}{\bibfnamefont{M.}~\bibnamefont{Lu}},
  \bibinfo{author}{\bibfnamefont{Y.-T.} \bibnamefont{Cui}},
  \bibinfo{author}{\bibfnamefont{M.}~\bibnamefont{Hashimoto}},
  \bibnamefont{et~al.}, \bibinfo{journal}{Phys. Rev. Lett.}
  \textbf{\bibinfo{volume}{110}}, \bibinfo{pages}{067003}
  (\bibinfo{year}{2013}).

\bibitem[{\citenamefont{Saito et~al.}(2013)\citenamefont{Saito, Onari, and
  Kontani}}]{Kontani_PRB:2013}
\bibinfo{author}{\bibfnamefont{T.}~\bibnamefont{Saito}},
  \bibinfo{author}{\bibfnamefont{S.}~\bibnamefont{Onari}}, \bibnamefont{and}
  \bibinfo{author}{\bibfnamefont{H.}~\bibnamefont{Kontani}},
  \bibinfo{journal}{Phys. Rev. B} \textbf{\bibinfo{volume}{88}},
  \bibinfo{pages}{045115} (\bibinfo{year}{2013}).

\bibitem[{\citenamefont{Jarrige}(2014)}]{Jarrige_Unpublished}
\bibinfo{author}{\bibfnamefont{I.}~\bibnamefont{Jarrige}},
  \bibinfo{journal}{Unpublished}  (\bibinfo{year}{2014}).

\end{thebibliography}
\end{document}